\documentclass[12pt,a4paper]{panl}
\usepackage{graphicx}
\usepackage{amssymb}
\usepackage{amsfonts}
\usepackage{amsmath}
\usepackage{longtable}
\usepackage{lscape}
\usepackage{epsfig}
\usepackage{multirow}

\newcommand{\ba}{\begin{eqnarray}}
\newcommand{\ea}{\end{eqnarray}}
\newcommand{\beqs}{\begin{eqnarray}}
\newcommand{\eeqs}{\end{eqnarray}}

\originalTeX
\begin{document}

\title{ Relations of meson and nucleon electromagnetic and gravitational radii with quarks and gluons contributions}
\maketitle

\maketitle
\authors{O.V.\ Selyugin$^{a}$\footnote{E-mail: selugin@theor.jinr.ru}
O.V. Teryaev$^{a,}$\footnote{E-mail: teryaev@theor.jinr.ru}
}
\setcounter{footnote}{0}
\from{$^{a}$\,BLTP JINR, Dubna, Russia}



\setcounter{footnote}{0}

\begin{abstract}
 The electromagnetic and gravitational 
 form factors
of the nucleon determined by quark and gluon contributions  are calculated using the momentum transfer dependence of generalized parton distributions with different forms of parton distribution functions obtained by various Collaborations.
  The power forms of gravitational form factors of quarks and gluons
  are examined.
It is shown that the gluon gravitational radius of the nucleon
  is comparable to the electromagnetic radius of the proton; however, the quark gravitational radius of the nucleon
  is less than its electromagnetic radius. It is shown
  that the gluon gravitational form factor drops faster than the quark
  gravitational form factor at large transfer momenta and corresponds to the triple form.

\end{abstract}
\vspace*{6pt}



  One of the central problems of hadron physics is the
  quark and gluon structure of  hadrons.
    The generalized parton distributions (GPDs)
 (which  depend on $\ x$, momentum transfer $t$,
  and the skewness parameter $\xi$)
 is one of the central
  physical approaches that allows one to obtain different hadron form factors from the same physical representation of the hadron structure.
 (see, for example \cite{Burk-N1,Pol-18,DD-N2,Ter-25}.
   It is tightly connected with spin physics of hadron \cite{Ter-gpd-sp}.

 One of the central parts of GPDs consists of the standard
 parton distribution functions (PDFs) of quark and gluons.
 They are extracted from different dip inelastic reactions 
 examined at HERA and LHC. 

   It was shown \cite{Mil94,Ji97,R97} that
the integration of  different  Mellin  moments
of $GPDs(x,\xi,t)$ 
 over $x$ gives us  different hadron form factors.
     This  is reflected as the the different Mellin moments over $x^{1-n}$.
     The first Mellin moment $n=1$ corresponds to the electromagnetic form factors of hadrons.
     When $n=0$, the GPDs give the Compton form factors and when $n=2$, one can obtain
     gravitomagnetic form factors of hadrons.

      Hence the GPDs unite a very wide region of experimental reactions.
      including elastic hadron scattering, where the electromagnetic   form factors
      play an essential role, and many different inelastic reactions,  
       used at HERA and LHC,
      from which PDFs were obtained.

    Now  the gravitational form factors of hadrons are  the most attractive.
    Different model approaches and lattice methods are used for the calculation of
    different gravitomagnetic form factors which enter
in the energy momentum tensor (EMT) (\cite{Pagels}).

%
 Since the GPD is not known a priori, one seeks for models of GPD
  based on general constraints on its analytic and asymptotic
 behavior. The calculated scattering amplitudes (cross sections) are then compared with the data to confirm, modify or reject the chosen form of the GPDs. \\



\vspace*{6pt}

 {\bf  
  Electromagnetic hadron form factors and transfer momentum dependence of GPDs} \\

 In \cite{GPD-ST-PRD09}
 the transfer momentum ($t$)-dependence of  GPDs  in the form
\ba
{\cal{H}}^{q} (x,t) \  &=& q(x)_{nf} \   exp [  a_{+}  \
  f_{q}(x) \ t ]
\label{GPD0}
\ea
 was researched  with $  f_{q}(x) = 
 (1-x)^2/x^{m}  $.
The function $q(x)$ was chosen at the same scale $\mu^2=1$ as in \cite{R04},
which is based on the MRST2002 global fit \cite{MRST02}.

   Further development of the model requires  careful analysis of the form
   of the momentum transfer of GPDs and the form of
   properly chosen  PDFs. In \cite{GPD-PRD14}, the analysis of  more than 24 different
   PDFs was made.
A complete set of the  existing experimental data on
    the electromagnetic form factors  was used .
  There were obtained
\ba
{\cal{H}}^{u} (x,t) \  = q(x)^{u} \   e^{2 a_{H}  \ f_{u}(x)
 \ t };  \ \ \  
{\cal{H_d}}^{d} (x,t) \  = q(x)^{d} \   e^{2 a_{H}
 f_{d}(x) \ t }.
\label{t-GPDs-H}
\ea
with $f_{u}(x) = 
  (1-x)^{2+\epsilon_{u}}(x_{0}+x)^{m} $ and
 $f_{d}(x) = (1+\epsilon_{0}) 
 (1-x)^{1+\epsilon_{d}}/(x_{0}+x)^{m}  $.

  The hadron
 form factors are related to the $GPDs(x,\xi,t)$
 by the 
 sum rules \cite{Ji97}
 \begin{eqnarray}
 F_1^q(t)=\int_{-1}^1 dx H^q(x,\xi=0,t),  \ \ \ \ F_1^q(t)=\int_{-1}^1 dx H^q(x,\xi=0,t).
 \end{eqnarray}

 The integration region can be reduced to positive values of
 $x,~0<x<1$ by the following combination of non-forward parton
 densities \cite{R04}
  ${\cal H}^q(x,t)=H^q(x,0,t)+H^q(-x,0,t)$,
   ${\cal E}^q(x,t)=E^q(x,0,t)+E^q(-x,0,t)$,
    providing
 $F^q_1(t)=\int_0^1 dx {\cal H}^q(x,t),\label{01}$,
 $F^q_2(t)=\int_0^1 dx {\cal E}^q(x,t).\label{02}$

 The proton and neutron Dirac form factors are defined as
 \begin{eqnarray}
 F_1^p(t)=e_uF_1^u(t)+e_dF_1^d(t),  \ \ \ \ F_1^n(t)=e_uF_1^d(t)+e_dF_1^u(t),
 \end{eqnarray}
  where $e_u=2/3$ and
 $e_d=-1/3$ are the relevant quark electric charges.
 As a result the $t$-dependence of the $GPDs(x,\xi=0,t)$
can be determined from the analysis of the nucleon form factors
for  which experimental data exist in a wide region of momentum transfer.
 It is a unique situation as it is unites the elastic and inelastic processes.

In the limit $t\rightarrow 0$ the functions $H^q(x,t)$ reduce to
 usual quark densities in the proton: $$ {\cal\
 H}^u(x,t=0)=u_v(x),\ \ \ {\cal H}^d(x,t=0)=d_v(x)$$ with the
 integrals $$\int_0^1 u_v(x)dx=2,\ \ \ \int_0^1 d_v(x)dx=1 $$
 normalized to the number of $u$ and $d$ valence quarks in the
 proton. \\

\vspace*{6pt}

     {\bf Gravitational form factors of nucleons} \\

   Taking the
 matrix elements of energy-momentum  tensor $T_{\mu \nu}$
 instead of
   the electromagnetic current $J^{\mu}$
    \cite{Pagels,Ji97,Pol-18}, we have  
 \begin{eqnarray}
 \left\langle p^{\prime}|\hat{T}^{Q,G}_{\mu \nu} (0)|p\right\rangle &=& \bar{u}(p^{\prime}) \biggl[
 A^{Q,G}(t) \frac{\gamma_{mu}P_{\nu} }{2}   
     + B^{Q,G}(t)\frac{i\left(P_{\mu}\sigma_{\nu \rho} + P_{\nu} \sigma_{\mu \rho}\right) \Delta^{\rho} }{4 M_N}  \nonumber \\ \nonumber
 &+& \ C^{Q.G}(t) \frac{  \Delta_{\mu} \Delta_{\nu} - g_{\mu \nu} \Delta^{2} }{M_N}
\biggr ] {u}(p)
\end{eqnarray}

The symmetric Belinfante-improved total QCD EMT operator is defined as
\begin{align}
    \hat{T}^{\mu \nu}=\sum_q T_q^{\mu \nu}+T_g^{\mu \nu}\ .
\end{align}
The quark and gluon contributions to the total EMT \cite{Cao-25} are given by
\begin{align}
    \begin{aligned}
    T_q^{\mu \nu}= & \bar{\psi}_q \left(\frac{i}{2}\gamma^{\{\mu}\mathcal{D}^{\nu\}}+\frac{1}{4}g^{\mu\nu}m_q\right)\psi_q\ , \\
    T_g^{\mu \nu}= & F^{A, \mu \eta} F^{A,}{ }_\eta{ }^\nu+\frac{1}{4} g^{\mu \nu} F^{A, \kappa \eta} F^{A,}{ }_{\kappa \eta}\ ,
\end{aligned}
\end{align}
where $\psi_q$ and $\mathcal{A}^A_\mu$ are the quark and gluon fields, respectively.
The total EMT is conserved; however, the gluon and quark parts of the EMT are not individually conserved.

 One can obtain the gravitational form  factors of quarks that are related to the second  moments of GPDs
\begin{eqnarray}
\int^{1}_{-1}dx \ x [H_q(x,\Delta^2,\xi)  = A^{q}(\Delta^2)+(-2\xi)^2C^{q}_{2,0}(\Delta^2).
\end{eqnarray}

 For $\xi=0 $ one has the second moment of GPDs
\ba
\int^{1}_{0} \ dx \ x \sum_{u,d}[{\cal{H}}(x,t) 
= A_{h}(t). 
\ea
The integration of the second moment of GPDs over $x$ gives  the momentum-transfer representation
  of the form factor. 
  It was  approximated  by the dipole form \cite{HEGS0} 
\begin{eqnarray}
  A(t)=A(0) \frac{L^{4}_{2}}{(L^{2}_{2}-t)^2}.   
 \end{eqnarray}

\vspace*{10pt}

 {\bf Nucleon mass  radii } \\

In practice, the nucleon GFFs are experimentally accessible because they are weighted integrals over the generalized parton distribution (GPD) functions, which can be accessed from hard exclusive processes such as deeply virtual Compton scattering~\cite{Ji97,R97} 
and hard exclusive meson production~\cite{Collins:1996fb}.
At from the theoretical side, various models (especially lattice calculations)
have been employed to explore the GFFs.
However, it should be noted that in most part these model-dependent results are typically subject to uncertainties that are difficult to control.

The mass radius  is defined as the radius derived from the energy density, see e.g.,
\cite{Ji-172}.
Lattice QCD (LQCD) was used to compute the quark contribution to the GFFs of the pion and proton.
The gluonic contribution to the GFFs is exanined for different hadrons~\cite{Pef-22,Shanahan-PRL,Wang-2024,Wang-2024-pi}.
A notable development is the recent LQCD calculations of the pion and nucleon GFFs~\cite{Hackett:2023nkr, Hackett:2023rif} with a pion mass of $m_\pi=170$ MeV, which is close to the physical pion mass.



%


 A good description of different form factors and  elastic scattering of  hadrons
 gives a good support to  our determination of the momentum transfer dependence of GPDs.
 Based on this determination of GPDs,
   let us calculate the gravitational radius of the nucleon using the integral representation of the form factor
  and make the numerical differentiation over $t$ as $t \rightarrow 0$. This method allows us
  to obtain a concrete form of the form factor by fitting the result of the integration of
   the GPDs over $x$.
   As a result, the gravitomagnetic radius is determined as
\begin{eqnarray}
<{r_{A}}^2> =  -\frac{6}{A(0)} \frac{dA(t)}{dt}|_{t=0};
\label{Rp}
\end{eqnarray}
hence the numerical derivative will be
\begin{eqnarray}
<{r_{A}}^2> =  -\frac{6}{A(0)} \frac{A(t_{1})-A(t_{1}+\Delta t)}{\Delta t};
\label{Rp}
\end{eqnarray}
   where
\begin{eqnarray}
A(t) = \int_{0}^{1} x \ 3 (q_{u}(x) + q_{d}(x)) \ e^{-\alpha \ t f(x)} dx.
\end{eqnarray}
 In the calculations of $A(t)$ PDFs were examined   
(in total more than 19 options)
 that were extracted from the experimental data by various   Collaborations \cite{GPD-PRD14}.
 In the numerical calculations of the nucleon mass radius 
the $t$ dependence of the starting point of the differentiation of the results, 
  $ -t_{1} =1. 10^{-3} \ \ \ {\rm and} \ \ \ \Delta t_{1} =1. 10^{-3} \ ;$
  $ -t_{1} =1. 10^{-2}  \ \ \ {\rm and} \ \ \   \Delta t_{1} =1. 10^{-3} \ ; $
  $ -t_{1} =4. 10^{-2} \ \ \  {\rm and}  \ \ \  \Delta t_{1} =1. 10^{-3}$  was compared.
 Finally, the arithmetical mean
 of different PDFs  was obtained for the minimal $t$:
$<r_{m}^{2}>^{1/2} = {\bf 0.54 \pm 0.02 \ fm \ }$ and for maximum examined $t$
$<r_{m}^{2}>^{1/2} = {\bf 0.52 \pm 0.02 \ fm }$.
 The comparison with the calculations of
 D.E. Kharzeev  \cite{Kharzeev-21}  
     $<r_{m}^{2}>^{1/2} = {\bf 0.55 \pm 0.03 \ fm }$.
 shows a good coincidence.
  It should be noted that the obtained matter radius of nucleon is essentially
  less than its charge radius
 $ <r_{C}^{2}>^{1/2} =  {\bf 0.8409 \ fm }$ \cite{Group-20}.

\vspace*{10pt}

  {\bf Meson gravitomagnetic form factors and radii } \\
The gravitational form factors of pions, kaons and the nucleons are investigated by employing modern dispersive techniques and chiral perturbation theory \cite{Shanahan-pi19,Cao-25}.
 They determine the gravitational form factors of pions and kaons, extending  analysis to explore the pion mass dependence of these form factors at several unphysical pion masses up to 391 MeV,

The tensor meson dominance 
model gives \cite{Donoghue} 
 $A^{\pi}(t) = m^{2}_{f_{2}} / (m^{2}_{f_{2}} -1) = 1 + t/m^{2}_{f_{2}}+ · · · $
$ m^{2}_{f_{2}} = (1275 \pm 20) $ MeV to cover all
for the
 $ m^{2}_{f_{2}}(1270)$ mass, namely $(1259 \pm 4 \pm 4)$ MeV  \cite{Navas,Donoghue}.



 So using meson PDF \cite{7} for the $\pi$-meson  we have
   $\chi^{2}/N = 0.26$ for the $\pi$-meson   which was obtained with $\Lambda^2 = 1.44$ and  $n=1.07$ .
 Hence the gravitational form factor of meson can be represented by
 the monopole form which leads to
  the mass radius of the $\pi$-meson equal to $0.67$ fm.

 The obtained value of the $\pi$-meson mass radius can be compared with the result
  $\sqrt{r^{2}_{\pi}} \ = \ 0.39 \ $ fm  \cite{Kumano-Ter}, in which it was noted
 that 
the charge radius $\sqrt{〈r^2〉} \ =  \ 0.672 \pm \ 0.008 \ $ fm for the
charged pion
is approximately $1.6$ times larger than
its mass radius. 
 In \cite{Hackett:2023nkr}, with calculations
from the lattice QCD
 monopole form pion gravitational form factors   were also  obtained     with
  $A^{\pi}_{q}(0) \ = \ 0.481(15) \ $ and $\Lambda= 1.26 \div 1.24 \ 1$   GeV$^2$;
and 
$A^{\pi}_{g}(0) \ = \  0.546(18)\ $ and $ \Lambda \ = \ 1.13 \div 1.11 \ $  GeV$^2$. \\

\vspace*{6pt}

 {\bf Gluon GPDs and gravitational radius} \\

  It is  important to determine the separate gluon contribution to
  the gravitational form factor and determine the gluon gravitational radius.
  We used six different gluon parton distributions $xg(x)$
  \cite{Broun2206.05672,BS-Bell,CTQE12,Vafaee,Bonvinib}.
Most of them have the polynomial form of $x$ and a more complicate form was obtained in \cite{CTQE12}.
   Especially note that $xg(x)$ was obtained in \cite{BS-Bell} 
    on the basis of
    the Bourrely-Soffer model and has a non-standard exponential form.
  As an example, we also used pion PDFs presented in work \cite{AriolaPi}.
  Further, we used the notation $u1,u2,u3,u4,u5,u6$ 
  in the order of references.

      Using the $t$ dependence of GPDs, obtained for the quarks contributions,
      the corresponding gluon gravitational form factors were obtained.
 \begin{eqnarray}
     A(t) = \int_{0}^{1} x g(x) \ e^{-\alpha \ t f(x)} dx
  \end{eqnarray}

      We found out that up to $|t| \leq 2$ their form is similar, but al large $t$ it essentially diverges.  Possibly, it is show that our picture for gluon distributions
    is  valid only for $|t| \leq 2$. To be consistent, we made a new fit for quark GPDs
       using only the experimental data in the region $|t| \leq 2$.
        Then we used a new parameter to obtained gluon gravitational form factors.
      Using these results the obtained gluon gravitational form factors are described by
      the multupole form with the fitting parameters $\Lambda$ and $n$.
   \begin{eqnarray}
     A_{g}(t) = A(0) \frac{\Lambda^2}{(\Lambda^2-t)^n}
  \end{eqnarray}
    The results were presented in Table 1.

\begin{table}
\label{Table-1}
\vspace{.5cm}
\begin{tabular}{|c|c||c|c||c|c|c|| } \hline
Npdf & ref    & $\chi^{2}_{tot}$   & $\Lambda^2$ GeV$^2$ & n & $A(0)$   &  $<r^2>^{1/2}$ fm     \\ \hline
  &    &   & f(x)=-Log(x)   &  &    &             \\ \hline
 u1  & \cite{Broun2206.05672} & 10  & 0.92  & 2.7    & 0.38  & 0.84       \\
 u2  & \cite{BS-Bell} & 183    & 0.93  & 3.    & 0.40  & 0.88    \\
 u3  & \cite{CTQE12} & 122   & 1.07  & 2.6    & 0.38  & 0.76    \\
 u4  & \cite{Vafaee} & 155300  &  0.65 & 3.     & 0.33  & 1.05       \\
 u5  & \cite{Bonvinib}$_{a}$ & 370    & 0.98    & 3.0   & 0.46  & 0.86      \\
 u6  &\cite{Bonvinib}$_{b}$  & 1380    & 0.96   & 3.     & 0.40  & 0.87        \\ \hline
\end{tabular}
\caption{Gluon gravitational radius corresponds to different gluon PDFs}
\end{table}
\vspace{.5cm}

\begin{figure*}
\includegraphics[width=.45\textwidth]{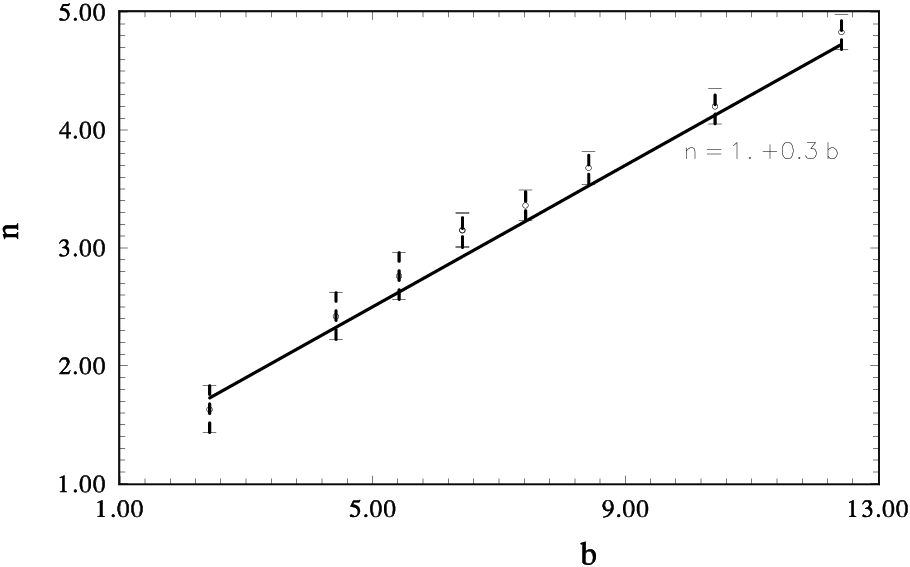}
\includegraphics[width=.45\textwidth]{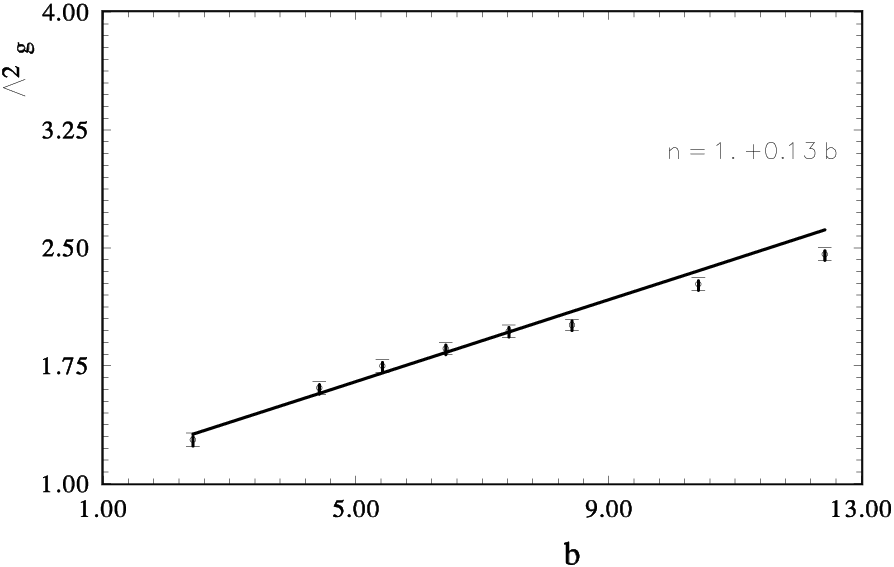}
\vspace{2cm}
\caption{a)[left]The dependence of $n_g$ of the gluon gravitational form factor
 on the parameter $b$;
b) [right] The dependence of $\Lambda^{2}_{g}$ of the gluon gravitational form factor
 on the parameter $b$ } 
\label{Fig_1}
\end{figure*}

\begin{figure*}
\includegraphics[width=.45\textwidth]{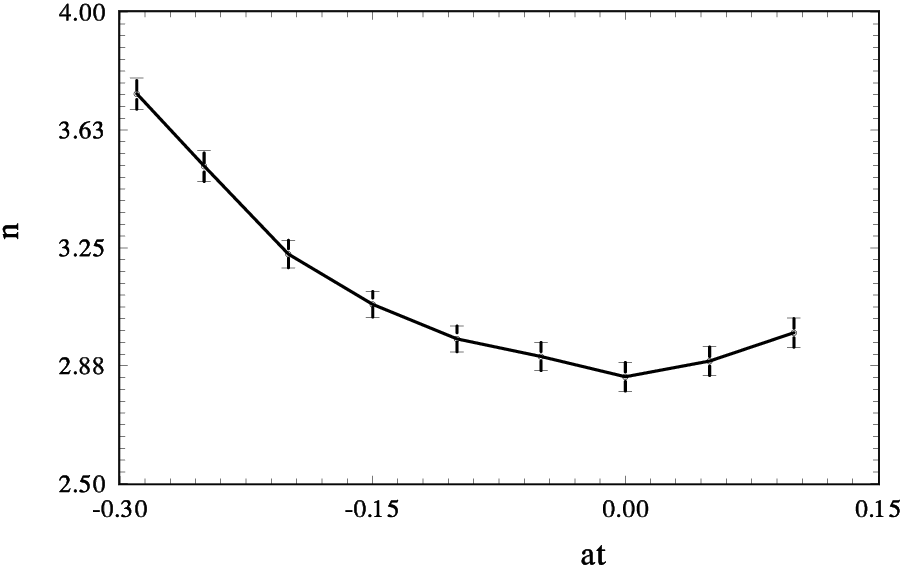} 
\vspace{2cm}
\caption{The dependence of $n_g$ of  gluon gravitational form factor over parameter $a$. }
\end{figure*}

   Our results show that there is a correlation between  the value of $n$ which is determined by
  the  dipole or triple form of the form factor and the value of $\Lambda^2$.
   Now let us examine how the matter radius of a hadron will  depend on
   the form of the gluon gravitational form factor. For this let us take that  $A(0)_g = A(0)_q =1/2$.
   Of course, it is some approximation but it is not essentially rough.
   In our model we have obtained that $A(0)_q = 0.54$.
   Some other models also give a close result,
    others obtained that $A(0)_q > 0.5$ but not so far from the middle value.
    Some of the quarks and gluons will be given for the matter radius
    \begin{eqnarray}
     <r^2>_{(q+g)} \  = \ \frac{1}{2} \ \frac{n_q}{\Lambda^2_{q}} \  + \  \frac{1}{2} \ \frac{n_g}{\Lambda^2_{g}}.
  \end{eqnarray}

   Now let us examine different cases: \\
   Case I)  (dipole -dipole case)\\
        Suppose that in both (quark and gluon) cases the gravitational
        form factors are described by  the dipole form. In this case, $n_q = n_g = 2$
        and for the radius we obtain
      \begin{eqnarray}
     <r^2>_{(q+g)}  \  = \ \frac{1}{2} \ \frac{2}{\Lambda^2_{q}} \  + \  \frac{1}{2} \ \frac{3}{\Lambda^2_{g}}.
\end{eqnarray}
  There can be  three different situations: \\
   a)  $\Lambda^2_{q} = \Lambda^2_{g}  $ \\
   This leads to $ <r^2>_{(q+g)} \  = \ <r^2>_{q} \ = \ <r^2>_{g} $ \\
  b)  $\Lambda^2_{q} \  = \ 1.5 \Lambda^2_{g}  $ \\
     This  leads that $ <r^2>_{(q+g)}  \  =   \frac{1.66}{ \Lambda^2_{g}}, \ $
    hence  $ \ <r^2>_{(q+g)}  \ $ is less than $ \ <r^2>_{g} \ $ \\
      but larger than $ \ <r^2>_{q} $  \\
    and $ <r^2>_{q} \  $ is less than $ \ <r^2>_{g}$
    but
    $ <r^2>_{(q+g)}  \ > \ \ <r^2>_{q} \ $ \\

   c)   $\Lambda^2_{g} \  = \ 1.5 \Lambda^2_{q}  $ \\
  In this case,
 $ \ <r^2>_{(q+g)}  \  =   \frac{1.65}{ \Lambda^2_{q}}, \ $
  or $=   \frac{2.5}{ \Lambda^2_{g}}, \ $
    $ <r^2>_{(q+g)}  \ \ < \ \  <r^2>_{q} \ $ \\

    Case II)  (dipole -tripole case)
    \begin{eqnarray}
     <r^2>_{(q+g)}  \  = \ \frac{1}{2} \ \frac{2}{\Lambda^2_{q}} \  + \  \frac{1}{2} \ \frac{3}{\Lambda^2_{g}}.
    \end{eqnarray}
   For a)  $\Lambda^2_{q} = \Lambda^2_{g}  $ \\
    we have  that $ <r^2>_{(q+g)}  \  =   \frac{2.5}{ \Lambda^2_{g}}$ \\
    and again we obtain that  $ <r^2>_{(q+g)}  \ $ is less $ \ <r^2>_{g} \ $
     but  large  $ \ <r^2>_{q} $  \\
     b)  $\Lambda^2_{q} \  = \ 1.5 \Lambda^2_{g}  $ \\
       This  leads that $ \ <r^2>_{(q+g)} \  =   \frac{2.16}{ \Lambda^2_{g}}$
     or $ = \frac{3.24}{ \Lambda^2_{q}}$.  \\
    Again,  this  leads that  $ <r^2>_{(q+g)} \ $ is less $ \ <r^2>_{g} \ $
    but   larger  $ \ <r^2>_{q} $  \\
    c)   $\Lambda^2_{g} \  = \ 1.5 \Lambda^2_{q}  $ \\
    then we obtain
    $  \ <r^2>_{(q+g)}  \  =   \frac{2.16}{ \Lambda^2_{g}}$,
    again  this  leads that  $ <r^2>_{(q+g)}  \ $ is less $ \ <r^2>_{g} \ $
     but  large than $ \ <r^2>_{q} $.

   The parameters $n_q$ and $\Lambda^{2}_{q}$ of quark gravitational form factor are determined sufficiently well.
   In our model, it is determined as $n_q = 2$ and  $ \Lambda^{2}_{q} = 1.58\pm 0.04$ and other model calculations
   are not so far from these values. On the contrary, for the gluon case, there is a large spread of the obtained values.
    Figure 1a shows that these values strongly  depend on the value of the parameter $b$
     and slightly depend on the parameter $a$.
     The coefficient $(1-x)^{b}$,
   determines PDFs (and 
    thus GPDs) as $x \rightarrow 1$ and therefore at large momenta.
   Our calculations show that in the gluon case, $A(0)_g$ is less than $0.5$.
    In most cases, it corresponds
  to $A(0)_{q} = 0.54$ obtained in our model.
 The size of the power value $n$ is near $3$  in most cases.
 The corresponding  sizes of $\Lambda^{2}_{g}$ are near $0.9\pm 0.2$
  which is essentially less than $\Lambda^{2}_{q}=1.6 \pm 0.1$.
  It  corresponds the case IIb)
  and $ \ <r^2>_{q} \  $ is less than $ \ <r^2>_{g}$.
  Hence  taking into account the gluon contributions
   increases the mass radius of the nucleon.

  If $n$ comes to $2$, the corresponding $\Lambda^{2}_{g}$
  grows essentially and exceeds the $\Lambda^{2}_{q}$.
   It corresponds to the case Ic).
  In this case,  $ \ <r^2>_{g} \  $ is less than $ \ <r^2>_{q}$
  and the gluon contributions
   decrease the mass radius of the nucleon.
  Such a situation contradicts the common model of nucleon
  that consists of the core of quarks surrounded by the gluon mezon cloud. \\

  \vspace*{6pt}

  {\bf Conclusions} \\

  Taking into account 
  different forms of quark and gluon PDFs obtained 
  by various Collaborations,  the generalized parton distributions are examined with
    a different form of $f(x,t)$ which determines the momentum transfer dependence of GPDs.
     This allow us to calculate the quark and gluons gravitational form factors.
     It was found that the quark gravitational mass form factor can be represented by the power  form -
$A(0)_{q} \Lambda^2/(\Lambda^2 -t)^{n}$ with $n=2$
  and with $\Lambda^{2}_{q} =1.6 \pm 0.1$.
 This is corresponds the dipole approximation.
  However, the gluon mass gravitational form factor is described
  by the power form with $n=3$, which corresponds to the triple approximation with $\Lambda^{2}_{g} =0.9 \pm 0.2$.
  The value of $A(0)_{q}$ is slightly above  half
 and the value of $A(0)_{g}$ is less than half.
 Their sum corresponds to the determination of
 $A(0)_{q} + A(0)_{g} \approx 1 $.
 Note that we do not take into account small additional contributions
  from the sea quarks and strange quarks.

 The dependence of the mass gravitational radius of nucleons corresponding to
  the sum of the quark and gluon gravitational mass radii
  was examined
  from the parameters of power representation.
  In Fig. 1 it is shown that the power $n$   
  depends a lot on the parameter $b$ and weakly depends on
  the parameter $a$  that determines the degree of dependence of the PDF
 on $x$.

It is shown that the gluon gravitational radius of the nucleon is comparable to the electromagnetic radius of the proton, however, the quark gravitational radius of the nucleon is less than its electromagnetic radius. It is shown that the gluon gravitational form factor drops faster than the quark gravitational form factor at large transfer momentum and corresponds to the triple form.

  For the meson (pion)  the momentum transfer dependence of GPDs  was taken the same as   for the nucleon.
   This made it possible to calculate
   the mass gravitational form factors of pion
    and use the calculated electromagnetic form factors for a
 good description of elastic pion-nucleon scattering at high energies.

   Our investigation of the nucleon structure shows that the density of the matter
   is more concentrated in the hadron than charge density and,
    correspondingly, the gravitational radius is less than the electromagnetic radius.
    The quark gravitational form factor $A_{q}(t)$ is represented by the dipole form
    with $\Lambda^2 = 1.6$ GeV$^2$.
    Using different gluon PDF,
    we obtain the gravitational form factor $A_{g}(t)$, which can be represented by the  multipole form
    with the power number $n=3$ and $\Lambda^2$ around $0.9$.
    Our analyses show that the  gluon contribution to the
    hadron gravitational form factor is approximately $45\%$
     and the quark contribution is $55\%$.
 The gluon gravitational radius obtained in our analysis
   appears to be  approximately equal to the electromagnetic radius.
   Therefore, gravitational form factor of the gluon and the corresponding
mass radius may be taken used to describe
the  coupling of pomeron to hadrons in hadronic interactions. This
approach would combine the correct C-parity of coupling with the current
phenomenological  success in the application of electromagnetic
form factor for this purpose.











%

%
 \end{document}